\documentclass[10pt,journal,compsoc]{IEEEtran}
\ifCLASSOPTIONcompsoc
  \usepackage[nocompress]{cite}
\else
  \usepackage{cite}
\fi
\ifCLASSINFOpdf
\else
\fi
\usepackage{amsmath}
\usepackage{amssymb}
\usepackage{mathtools} 
\usepackage{xcolor}
\usepackage{comment}
\usepackage{ragged2e}
\usepackage{makecell, caption}

\begin{document}
%
\title{Dynamic multi feature-class Gaussian process models}
%
%
%
%

\author{Jean-Rassaire Fouefack,
        Bhushan Borotikar,
        Marcel L\"uthi,
        Tania S. Douglas, 
        Val\'erie Burdin,
        and~Tinashe~ E.M.~Mutsvangwa,~
\IEEEcompsocitemizethanks{\IEEEcompsocthanksitem Jean-Rassaire Fouefack, Tania Douglas and Tinashe Mutsvangwa are with the Division of Biomedical Engineering,
        University of Cape Town, 7935, South Africa.\protect\\
E-mail: fjrassaire@gmail.com,\protect\\ tinashe.mutsvangwa@uct.ac.za
\IEEEcompsocthanksitem Jean-Rassaire Fouefack, Val\'erie Burdin  are  with the IMT-Atlantique and the Laboratory of Medical Information Processing (LaTIM INSERM U1101), Brest, France.\protect\\
E-mail: valerie.burdin@imt-atlantique.fr
\IEEEcompsocthanksitem Bhushan Borotikar is  with the Symbiosis Centre for Medical Image Analysis, Symbiosis International University, Pune, India.
\protect\\
E-mail: bhushan.borotikar@gmail.com 
\IEEEcompsocthanksitem Marcel L\"uthi, is  with , University of Basel, Basel 4001, Switzerland\protect\\
}
\thanks{This work is based on research supported by the National Research Foundation (NRF) of South Africa (grant no's 105950 and 114393); the South African Research Chairs Initiative of the NRF and the Department of Science and Technology (grant no 98788); and Brest M\'etropole (Mission Enseignement Sup\'erieur, Recherche et Innovation, grant MNB/MJM no 17-178), France}}

\IEEEtitleabstractindextext{%
\begin{abstract}
\justifying
In model-based medical image analysis, three features of interest are the shape of structures of interest, their relative pose, and image intensity profiles representative of some physical property. Often, these are modelled separately through statistical models by decomposing the object's features into a set of basis functions through principal geodesic analysis or principal component analysis. However, analysing multiple objects in an image using multiple single object models may lead to large errors and uncertainties, especially around organ boundaries. A question that comes to mind is what kind of advantages can be gained from combining the three features of interest in the same statistical space for analysing images. This study presents a statistical modelling method for automatic learning of shape, pose and intensity features in medical images which we call the Dynamic multi feature-class Gaussian process models (DMFC-GPM). A DMFC-GPM is a Gaussian process (GP)-based model with a shared latent space that encodes linear and non-linear variation. Our method is defined in a continuous domain with a principled way to represent shape, pose and intensity feature classes in a linear space, based on deformation fields. A deformation field-based metric is adapted in the method for modelling shape and intensity feature variation as well as for comparing rigid transformations (pose). Moreover, DMFC-GPMs inherit properties intrinsic to GPs including marginalisation and regression. Furthermore, they allow for adding additional pose feature variability on top of those obtained from the image acquisition process; what we term as permutation modelling. For image analysis tasks using DMFC-GPMs, we adapt Metropolis-Hastings algorithms making the prediction of features fully probabilistic. We validate the method using controlled synthetic data and we perform experiments on bone structures from CT images of the shoulder to illustrate the efficacy of the model for pose and shape feature prediction. The model performance results suggest that this new modelling paradigm is robust, accurate, accessible, and has potential applications in a multitude of scenarios including the management of musculoskeletal disorders, clinical decision making and image processing.
\end{abstract}

\begin{IEEEkeywords}
 Shape-pose-intensity latent space, Gaussian process, Shoulder joint analysis.
\end{IEEEkeywords}}

\maketitle

\IEEEdisplaynontitleabstractindextext

%
\IEEEpeerreviewmaketitle

\IEEEraisesectionheading{\section{Introduction}\label{sec:introduction}}

%
%
%
%

\IEEEPARstart{O}{ver} the past decades, model-based strategies for medical image analysis have been popular for the automatic extraction of quantitative information from images \cite{rueckert2019model}. Typically, these techniques analyse anatomical objects in images by decomposing the object's features into a set of basic functions (latent variable), which parameterise the distribution of the object's features in their natural space. In medical images, these features have three main classes: shape, pose and intensity. Although these are different types of features, coming from different distributions and features spaces, they may be correlated \cite{breton2014study,cootes1998active}. Indeed, joint poses are influenced by the shape of their articular geometries, which contributes to the regulation of their relative position. It also makes sense to consider an object's boundary in a medical image  to be implicitly defined by contrast differences relative to its surroundings. Regardless of these correlations, to date, advanced model-based algorithms in medical imaging have largely focused on each of these features shape, pose and intensity, separately  \cite{rueckert2019model}. These models include: statistical shape models (SSMs) or point distribution models (PDMs) \cite{cootes1992training} \textbf{\textit{for shape modelling}}; statistical shape and pose models (SSPMs) \cite{bossa2007multi} \textbf{\textit{for pose modelling}}; and  active shape models (ASMs) \cite{cootes1995active} \textbf{\textit{for intensity modelling}}. These models work by statistically analysing objects' features using principal component analysis (PCA) or principal geodesic analysis (PGA). In the PCA-based models, variability across features is assumed to be linear (shape and intensity variability),  while in the PGA-based models, such variability is considered to be non-linear (pose variability). The state-of-the-art in shape modelling are Gaussian process morphable models (GPMMs) \cite{luthi2017gaussian} which extend PDM-based shape variation modelling to a continuous domain. However, similarly to PDMs, GPMMs are PCA-based and thus assume linearity in variation. This makes them unsuitable for encoding pose variation which inherently incorporates rotation of objects; a non-linear distribution of transformations. Additionally, the Gaussian processes (GPs) in GPMMs are computed as geometry-based deformation fields, precluding them from being used directly for non-geometric feature modelling such as would be required for intensity modelling. Statistical shape and pose model approaches incorporate the modelling of pose variation into PDMs. However, current approaches utilise standard pose representation using Rodrigues, quaternion or Euler angles which often results in large pose prediction errors, especially when applied to non-compact objects. Statistical shape models are often used to perform segmentation through ASMs \cite{cootes1995active} where the objects are segmented through simultaneous use of the SSMs and an average intensity profile of the boundary of objects. However, the average intensity profile may not be sufficient to delineate the object boundary. Furthermore, such a process does not provide access to the full internal intensity profile of objects, instead focusing only on the boundary delineation. This is a missed opportunity as such intensity information maybe a clinically useful, for example, calibrated image intensity in CT or X-ray is useful proxy for understanding bone density or bone quality. 
 \begin{figure*}[ht]
	\centering
	\includegraphics[width=1.0 \textwidth, angle =0 ]{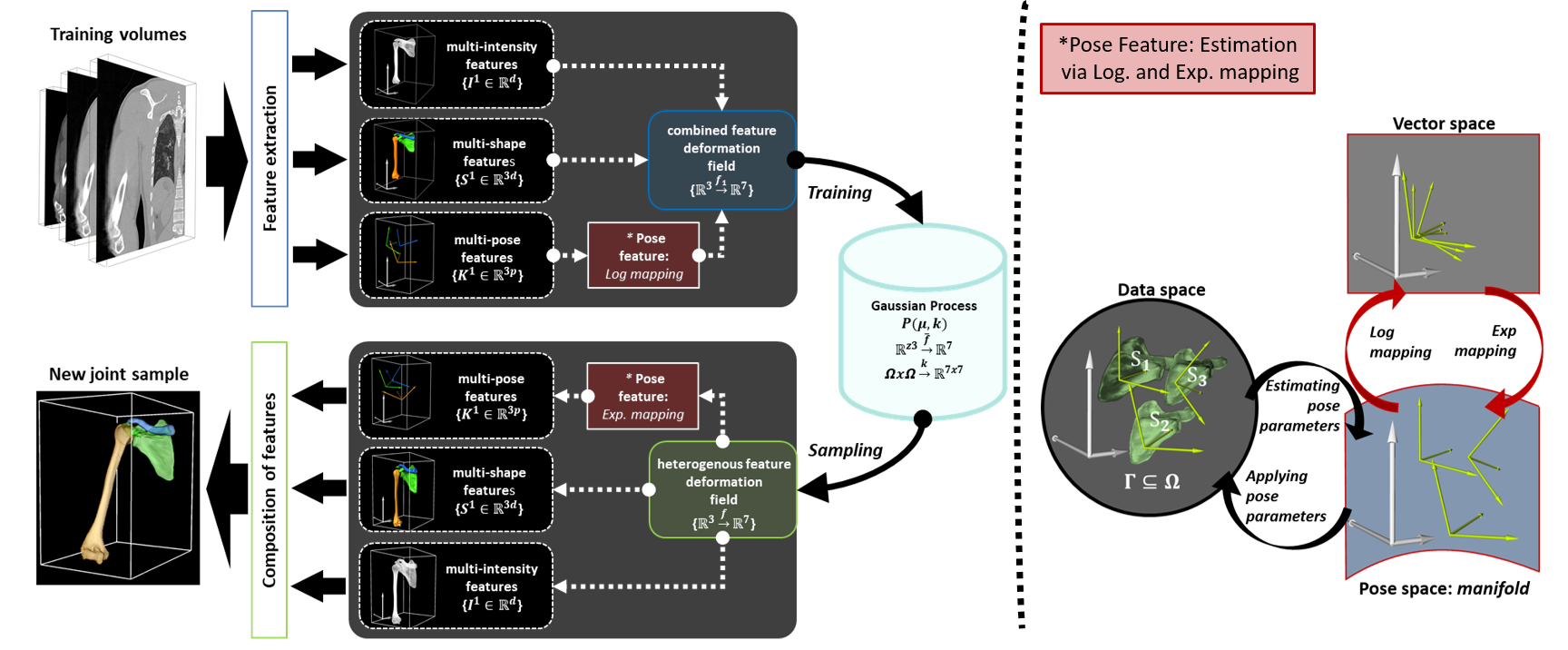}
	\caption[Proposed multi feature-class models architecture]{Proposed multi feature-class model architecture. Left: Shape, pose and intensity features are extracted from the training volumes. From these features, combined deformation fields are defined for model training by computing the Gaussian process.  New joint samples are generated from the model by sampling heterogeneous deformation fields. Right: computation of the logarithmic and exponential mappings for pose features.  The rigid transformations belonging to the manifold are obtained by estimating the pose parameters. These rigid transformations are projected into the vector space using logarithmic mapping to obtain linearised pose features. The reverse is done through exponential mapping when sampling from the latent space.} 
	\label{Globalframework}
\end{figure*}
 \subsection{Contributions}
 In this paper we propose a novel approach for encoding the variation of multiple feature classes in the same statistical space. We focus on modelling the shape and intensity of multiple objects in medical images together with their relative spatial orientations resulting in what we term  \textbf{\textit{dynamic multi-feature class Gaussian process models (DMFC-GPM)}}\footnote{A tutorial is available online for research use: https://rassaire.github.io/dmo-gpm-tutorial/ \label{tutorial}}. This approach extends GPMMs, SSPMs and ASMs by defining a GP-based shared latent space of shape, pose and intensity in a continuous domain while appropriately managing the non-linearity of the pose variation through the use of PGA. Deformation field-based representation of the pose is used instead of standard pose representation which is not robust when comparing non compact objects due to the object elongation effect \cite{moreau2017new}. The incorporation of all the feature classes (shape, pose and intensity) allows for the extraction of quantitative information of multiple objects of interest present in a medical image.
 
 Furthermore, the modelling approach inherits the intrinsic properties of GPMMs, which include on-demand increase of model flexibility, model marginalisation, and regression. However, while GPMMs focus on shape only, our modelling approach extends the utility of these intrinsic properties to pose and intensity modelling. Concretely, our modelling approach extends: 1) the ability to increase the variability around the mean for all three features of the model without retraining; 2) the ability to marginalise over a subdomain, to all features: from a specific class, from a specific structure or from different classes. In addition to the region of interest shape marginalisation allowed by GPMMs, the DMFC-GPM approach goes beyond shape and allows single shape, multiple shape-pose, intensity or region based multiple feature marginalisation; and 3) extends regression to any of the features included in the domain. One of the practical applications of this regression is to constrain a model at the desired relative pose. Our approach allows for developing a more flexible model using limited poses in training data, leading to what we call a permutation model. Permutation modelling may allow for prediction of permissible motion around a joint from single image acquisition of each of the joints in the training data. This would avoid the need for multiple pseudo-static 3D image acquisitions in 3D image based motion analysis studies, reduce patient radiation if imaging is radiation dependent, and allow analysis in cases where there is a truncated field of view (FOV) of the joint of interest.
 
By combining our modelling approach, the DMFC-GPM, with a model fitting approach, we introduce the DMFC-GPM framework (in this paper "framework" refers to modelling \textit{plus} model fitting). For model fitting we propose a customised Metropolis-Hastings algorithm adopted to fit DMFC-GPM models to observations through sampling. In addition, we show how the incorporation of a projection method in the model sampling process allows for sampling from the $3D$ latent space directly to $2D$. Figure \ref{Globalframework} shows the flowchart of the proposed model together with a manner to sample from the model. 

The DMFC-GPM framework is critically evaluated and validated through controlled experiments and is also applied to the problem of  analysing shape, pose and intensity features in medical images. We illustrate the importance of having a shared latent space of shape, pose and intensity feature classes. In particular, we show how correlations between feature classes can be leveraged for prediction, and how the leveraging of correlations between object shapes can improve segmentation accuracy. Furthermore, we apply the framework to analyse shoulder joints through first, estimating a pre-morbid shape and pose from a partial observation of the joint. Here we show that the incorporation of shape and pose features can provide additional information for estimating each of the bones in a joint, independent of spatial orientation. Second, we predict joints from $3D$ images, and show that the DMFC-GPM framework may be useful in determining the planning target joint anatomy and kinematic information from the available geometric features and $3D/2D$ images.

The contributions of this work are:
\begin{itemize}
    \item A principled way to represent shape, pose and intensity feature classes in a linear space, based on deformation fields.  
    \item Adaptation of a deformation field-based metric for comparing rigid transformations.
    \item A Gaussian process based latent space for multi feature-class modelling.
    \item Application of the modelling framework to real human shoulder computed tomography (CT) data.
\end{itemize}

\subsection{Related work}\label{Previous work} 
This section provides a brief literature review on shape, pose and intensity models over which our modelling approach is built. We synthesise the current state-of-the-art in statistical modelling of shape, pose and image-intensity along with their limitations. 

\textit{Statistical modelling of shape}: Regarding analysis of shape, a well established and understood formalism for analysing $3D$ geometric variation in a linearized statistical space exists in the form of SSMs. Statistical shape models typically model the data in a Euclidean vector space using PCA that treats feature variation as a linear combination of local displacements only \cite{cootes1992training,blanc2012confidence,mutsvangwa2015automated}. While efforts for faithfully representing the non-linearity of  shape space have been reported \cite{von2018efficient}, they have not become mainstream due to computational inefficiency and a lack of robustness. In addition, the validity of linearising shape space for rigid shapes has been codified in the literature for single anatomical structures \cite{cootes1992training,blanc2012confidence}. 

Recently, \cite{luthi2017gaussian} introduced a generalisation of SSMs, referred to as GPMMs. In contrast to discrete models (for example, PDMs) that are dependent on artificial discretization, GPMMs are inherently continuous, that is, permitting of the arbitrary discretization of the domain on which the model is defined. In our approach, the parametric low-dimensional model is represented as a GP over deformation fields obtained from training examples, adopting the same concept of shape modelling as in GPMM. In contrast to GPMMs, we consider objects of interest in their physiological context, that is considering the articulated anatomical complexes composed of several rigid substructures. This allows the integration of inter-object shape correlations in the latent space, which is important for understanding anatomical-physiological relationships between articulated objects. 

\textit{Statistical modelling of shape and pose}: Similar to the shape case, pose variation analysis using linear descriptors has been previously reported \cite{smoger2015statistical,klinder2008spine,fitzpatrick2011development,agrawal2020combined}. Reports abound in the literature on efforts to use methods more suited for managing the non-linearity of pose descriptors \cite{anas2014statistical,bossa2007multi,fletcher2004principal,chen2014automatic}. These approaches use parametric low-dimensional models of several subjects in different poses with the pose variation representation through PGA \cite{fletcher2004principal,bossa2007multi}. Additionally, shape and pose PGA-based models encountered in the literature use a standard representation (SR) for modelling the rotations describing the pose, that is, Rodrigues, quaternion and Euler representations. Rather than using SR, our approach leverages a deformation-based representation similar to \cite{moreau2017new}, but with resting position retrieval of the object. This avoids incorrect representation of translation due to the object's shape. 

\textit{Statistical modelling of intensity}: In addition to shape and pose features, another feature class from the raw images, namely intensity or grey-level distribution, has been considered in the literature. Starting from \cite{cootes2001active,cootes1992training}, statistical models of appearance have been applied to images using a set of model parameters for both shape and intensity variation learnt from training image data. The same idea has been used on various SSPMs in the literature to predict structures from the observed images \cite{anas2016automatic,sebastian2003segmentation,klima2016intensity,fotsin2019shape,delingette2021}. Typically, the approach is to simultaneously use shape-pose priors and controllable intensity models in segmentation/registration algorithms. However, none of these models embeds the grey-level features in the shape-pose priors. Additionally, these methods require a discretization of the structures of interest; structures which are continuous by nature. In our framework, intensity is modelled using deformation fields in a continuous domain and integrated in a shared latent space together with the pose and shape features. As we will show later in this paper we believe this formulation leads to more general analysis applications than previous frameworks for model based analysis. For example, our framework can be used for object segmentation, object completion, pose prediction, and  reconstruction of $3D$ from $2D$. 

To summarise, the idea of embedding all feature classes (shape, pose, intensity) in a single and continuous model is not yet explored in the literature.  This paper extends previous work for modelling the shape and pose of multiple objects to volumetric modelling of the shape and intensity of multiple objects with relative spatial orientation in medical images \cite{fouefack2019statistical,fouefack2020dynamic}. The framework brings new affordances for multi-object segmentation, $3D$ and $2D$ reconstruction, registration, and object recognition. The rest of the paper is organised as follows: The DMFC-GPM is presented in section \ref{DMFC-GPMs}. Next we present some interesting properties of the modelling approach in section \ref{Properties}. Section \ref {Experiments and Results} adds a model fitting method resulting in the DMFC-GPM framework. Experiments to validate the framework are then presented including benchmark comparisons against other commonly used approaches. We apply the framework to real image data of the shoulder in section \ref{applications}. The paper ends with discussion of the results and the full framework section \ref{Discussion}, with conclusions drawn in Section \ref{Conclusion}.

\section{Dynamic multi feature-class Gaussian process models}\label{DMFC-GPMs}
\subsection{Single object models}
 We consider an object, for example a bone in a medical image, to be characterised by three classes, which are the shape (S), pose (P) and the intensity (I). The shape indicates the surface of the object. The pose feature is an arbitrary position of the object relative to surrounding structures during image acquisition. The intensity is the grey-level profile inside the object. The shape, pose and intensity deformation fields are the features indicating variation across the training data for each of the classes. Let us consider the reference $\Gamma$, lying in $\Omega$, a subset domain of $\mathbb{R}^3$. By considering a shape feature as a vector in $\mathbb{R}^3$, a pose feature as a vector in $\mathbb{R}^3$, and an intensity feature as a real number, we define an object $V$ with respect to the reference $\Gamma$, as a function $f \colon \Omega\longrightarrow \mathbb{R}^3\times\mathbb{R}^3\times\mathbb{R}$ defined in the domain $\Omega\subset \mathbb{R}^3$. Thus,  $f$ maps each element of the domain to a septuplet $(\mathcal{S}, \mathcal{P}, \mathcal{I})$. Here $\Omega$ contains the reference $\Gamma$, and  the septuplet concatenates shape, pose and intensity feature vectors. 
 
Because each feature class variation (shape, pose or intensity) belongs to its own space, different by nature, units, and distribution (the amplitude of the variability across the dataset), we need to compute class specific deformation fields while taking into account the properties of their particular spaces. As detailed below, we factorise (categorise)  these features, into linear space for shape and intensity features, and into non-linear space for pose features. 

\subsubsection{Shape and intensity functions}
In the DMFC-GPM approach, we assume $n$ training examples and that the shape and intensity belong to a vector space. It is important to note that these $n$ learning examples are also assumed to be in correspondence, i.e. there is a one-to-one correspondence between the shape and intensity features across them. The shape space is obtained using generalised Procrustes analysis (GPA) \cite{cootes1992training,gower1975generalized} after obtaining correspondence. For the computation of the deformation field $\mathcal{S}_i$, the GPMM approach \cite{luthi2017gaussian} is used. The  deformation field $\mathcal{S}_i$ maps the reference $\Gamma$ to the $i^{th}$ object in the shape space (i.e. the one obtained after GPA) and we define it to be:
$ \mathcal{S}_i:\Omega\longrightarrow\mathbb{R}^{3},\:
    x\longmapsto \mathcal{S}_i(x)$. The mean shape deformation field is defined as:
$\displaystyle{
\overline{\mathcal{S}}\; \colon\Omega\longrightarrow\mathbb{R}^{3},\:
    x\longmapsto \frac{1}{n}\sum_{i=1}^n\mathcal{S}_i(x) }$
    
The deformation field $\mathcal{I}_i$ maps the reference intensity profile of $\Gamma$ to the target and we define it as:
$ \mathcal{I}_i:\Omega\longrightarrow\mathbb{R},\:
    x\longmapsto \mathcal{I}_i(x)$ with the mean intensity deformation field defined as: 
$\displaystyle{
\overline{\mathcal{I}}\; \colon \Omega\longrightarrow\mathbb{R},\:
    x\longmapsto \frac{1}{n}\sum_{i=1}^n\mathcal{I}_i(x)
}$
 
\subsubsection{Pose feature function}
Pose features are defined as pose variation of an object relative to the reference $\Gamma$. Contrary to shape and intensity, the pose of an object is represented by a rigid transformation. Rigid transforms belong to a manifold rather than Euclidean space and have to be analysed using non-linear ordination methods. Similarly to \cite{moreau2017new} we represent our pose variation as a deformation field leading to $\mathcal{P}_i$.  However, when estimating rigid transformations that include rotation between the reference and the target objects, we need to be careful as the invariant point through a rotation is the centre of mass, whereas the invariant point of motion in a rotating body is the rotation centre. We circumvent this problem by composing the estimated rigid transformation with a translation $\mathcal{T}_i$, thus making the centre/axis of rotation invariant in rotation. This allows us to get the neutral position (rest pose) of the object. This transformation representation, which we call as energy displacement representation, will be referred to as EDR for the remainder of the paper. In addition, for notation, an arrow on top of symbol represents sets of point clouds transformed into a single vector. The pose feature function $\mathcal{P}_i$ is then defined as:

\begin{align}
\label{EDR}
\mathcal{P}_i=\log[h_i]  \nonumber \\
\log[h_i](x)=h^{-1}_i\circ \mathcal{T}_i(x)-Id(x), x \in  \Omega \nonumber \\
h_i=\underset{ h\in SE(3)}{\arg min} \left \Vert \overrightarrow{h(V_i)}-\vec{\Gamma}\right \Vert ^2
\end{align}

The resulting Euclidean norm (non-linear metric) associated the rigid transformation space ($\textrm{SE}(3)$) is defined as : 
\begin{align}\label{EDR norm}
 d^2(h_i,Id)=\overrightarrow{\mathcal{P}_i( \Gamma)}^T\overrightarrow{\mathcal{P}_i( \Gamma)}
\end{align}

The Fr\'echet mean of pose transformation $\overline{\mathcal{P}}$, is directly obtained from the mean deformation fields, hence, its computation is more efficient.  The mean pose $\overline{\mathcal{P}}$ can now be defined by: 

\begin{align}
\overline{\mathcal{P}}(x)=\bar{h}^{-1}\circ\overline{\mathcal{T}}(x)-Id(x),x \in  \Omega \nonumber \\
\text{with}\; \bar{h}=\underset{h\in \textrm{SE}(3)}{\arg min} \;\left \Vert \overrightarrow{h\left [\frac{1}{n}\sum_{i=1}^{n}\mathcal{P}_i(\Gamma)\right ]}-\vec{\Gamma}\right \Vert ^2  \nonumber \\
\text{and}\; \overline{\mathcal{T}}(x)=\frac{1}{n}\sum_{i=1}^{n}\mathcal{T}_i(x)
\end{align}

\subsubsection{Building the Gaussian process models}\label{Computing training functions}
We assume that the shared latent space combining shape, pose and intensity variation features can be learnt from a set of training functions $\{f_i\}_{i\in \mathbb{N}}$, which are defined from the set of objects $\{V_i\}_{i\in \mathbb{N}}$. The function $f_i$ maps the domain to septuplet combining variation of shape, pose and intensity features. Let us denote the function that maps an element $x$ of the domain to its $i^{th}$ target $f_i(x)$ combining shape, pose and intensity. The function $f_i$ is then:

\begin{align}\label{training deformation}
 f_i\; \colon\Omega &\longrightarrow
  \mathbb{R}^{3}\times\mathbb{R}^{3}\times\mathbb{R} \nonumber \\
    x&\longmapsto  f_i(x)=\left(\mathcal{S}_i(x),\mathcal{P}_i(x),\mathcal{I}_i(x)\right)
\end{align}

The mean function $\bar{f}$ can now be defined by:

\begin{align}\label{mean deformation}
\bar{f}\colon\Omega \longrightarrow &\mathbb{R}^{3}\times\mathbb{R}^{3}\times\mathbb{R}  \nonumber\\
    x \longmapsto & \left(\overline{\mathcal{S}},\overline{\mathcal{P}},\overline{\mathcal{I}}\right)
\end{align}

with the  sample kernel ${\mathcal{K}}$ being: 

\begin{align} \label{kernel}
\mathcal{K}(x,y)=\frac{1}{n-1}\sum_{i=1}^{n}(f_i(x)-\bar{f}(x))(f_i(y)-\bar{f}(y))^T, \nonumber \\ 
(x, y)\in \Omega \times \Omega
\end{align}

where $n$ is number of training examples; $\overline{\mathcal{S}}$, $\overline{\mathcal{P}}$, and $\overline{\mathcal{I}}$ are the mean shape, pose and intensity deformation fields, respectively.

Then, we compute the shared latent space as a Gaussian process 
$f\sim \mathcal{GP}(\bar{f}, \mathcal{K})$ with $\bar{f}$ being the mean function and $\mathcal{K}: \Omega\times \Omega\longrightarrow (\mathbb{R}^3\times\mathbb{R}^3\times\mathbb{R})^2$ being the sample kernel function. The Gaussian process $\mathcal{GP}$ is parameterized by a set of basis functions $\{\Phi_m\}_{m\in \mathbb{N}}$, and its generative representation is then:
\begin{align}\label{DMFC gp}
f_{\theta}(x) \sim \bar{f}(x)+\sum_{m\in \mathbb{N}}\theta_m\sqrt{\lambda_{m}}\, \Phi_{m}(x), x\in \Omega
\end{align}

where $(\lambda_m, \Phi_m: \Omega\longrightarrow \mathbb{R}^3\times\mathbb{R}^3\times\mathbb{R})$ are the pair of eigenvalues and eigenfunctions similar to those of the integrator operator in \cite{luthi2017gaussian}, and $\theta=(\theta_m)_m \sim N(0,1)$ a Gaussian noise. The Gaussian process is made continuous by interpolating the functions $\{f_i\}_i$ as in \cite{meyer2002generalized} and  \cite{luthi2017gaussian}.

\subsection{Generalising to multi-object models}
We have derived the formulation of our models (Eq. \ref{training deformation}) for the shape, pose and intensity of an object in a continuous domain. However, in practice, pose is usually of interest when we want to analyse several articulating objects; for example, the humerus, scapula and clavicle which constitute the shoulder complex. We now generalise the framework to a finite number of objects. 

Let now assume that a training example ($V_i$) is a joint with $N$ objects. We defined $V_i$ as:
\[V_i=\Big\{V_i^1,\dots,V_i^N\Big\}\]

The reference joint $\Gamma$ is defined as the union of all $j$ individual references, that is $\Gamma= \bigcup_{j=1}^{N}\Gamma^j\subset\Omega=\bigcup_{j=1}^{N}\Omega^j$ with $\Omega^j$ being the $j^{th}$ object's domain.

The function $f_i$ representing the joint object $V_i$, is defined as
\begin{align}\label{mean deformation-multi}
  f_i\colon \cup_{j=1}^{N}\Omega^j &\longrightarrow\mathbb{R}^{3}\times\mathbb{R}^{3}\times\mathbb{R} \nonumber  \\
    x& \longmapsto f_i(x)=f_i^j(x), \text{ if } x\in \Omega^j
\end{align}
with $f_i^j(x)=\left (\mathcal{S}^j_i(x),\mathcal{P}^j_i(x),\mathcal{I}^j_i(x)\right )$.
The deformation field $f^j_i$ maps the reference to the $j^{th}$ object in the $i^{th}$  training joint example.  The shape $\mathcal{S}^j_i$ deformation maps the $j^{th}$ reference shape to the $j^{th}$ target shape in the $i^{th}$ joint,  $\mathcal{P}^j_i$ rigidly transforms the $j^{th}$ reference to the $j^{th}$ object position and $\mathcal{I}^j_i$  maps the $j^{th}$ reference intensity  to the $j^{th}$ target intensity in the $i^{th}$ joint.

\subsection{Computing new objects from the model}
After the parameterization of the shared latent space (Eq. \ref{DMFC gp}),  we need to develop a method to sample from the shared latent space. For a given Gaussian noise $\theta$, the shared sampled function $f_{\theta}$ maps a point $x$ in the domain $\Omega$ to a triple vector containing the shape, pose and intensity components. As for the definition of $f_i$ (Eq. \ref{training deformation}), the shape component is defined by the deformation field $\mathcal{S}_{\theta}$, the pose component by the deformation field (rigid transformation) $\mathcal{P}_{\theta}$, and the intensity component by the deformation field $\mathcal{I}_{\theta}$. The function $f_{\theta}$ is defined as:

\begin{align}\label{new deformation}
  f_{\theta}\colon\Omega&\longrightarrow\mathbb{R}^{3}\times\mathbb{R}^{3}\times\mathbb{R}  \nonumber \\
    x&\longmapsto f_{\theta}(x)=\left( \mathcal{S}_{\theta}(x),\mathcal{P}_{\theta}(x),\mathcal{I}_{\theta}(x)\right)
\end{align}

\subsubsection{Computing new 3D samples}
A new object is obtained from three steps. First, we deform the reference using the new shape deformation field; second, we apply the new pose deformation field (rigid transformation) to obtain its corresponding spatial orientation; third, we obtain the corresponding intensity by warping the reference intensity using the intensity deformation field. This is succinctly computed using:   
\[V_{\theta}=\left \{\left(\mathcal{P}_{\theta}\circ \mathcal{S}_{\theta}(x),\mathcal{I}_{\theta}(x)\right)| x\in \Omega\right \}\]

Let us now explicitly define the deformation fields  $\mathcal{S}_{\theta},\mathcal{P}_{\theta}$ and $\mathcal{I}_{\theta}$. To do so, the shared basic function $\Phi_m$ (Eq.\ref{DMFC gp}) needs to be defined. Similar to $f_{\theta}$, $\Phi_m$ is defined with shape ($\mathcal{S}_{\Phi_m}$), pose ($\mathcal{P}_{\Phi_m}$) and intensity ($\mathcal{I}_{\Phi_m}$) components as: 

\begin{align}\label{basic function}
  \Phi_m\colon&\Omega\longrightarrow\mathbb{R}^{3}\times\mathbb{R}^{3}\times\mathbb{R}  \nonumber \\
    &x\longmapsto \Phi_m(x)=\left(\mathcal{S}_{\Phi_m}(x),\mathcal{P}_{\Phi_m}(x),\mathcal{I}_{\Phi_m}(x)\right)
\end{align}

As mentioned in section \ref{DMFC-GPMs}, the shape and intensity deformation fields are linear transformations, so the new shape and intensity deformation fields of the shared latent space are defined as a linear combination of the parametric function and the eigenvalues. Thus:
\[\mathcal{S}_{\theta}(x)=\overline{\mathcal{S}}(x)+\sum_{m=1}^{M}\theta_m\sqrt{\lambda_{m}}\mathcal{S}_{\Phi_m}(x)\]
\[\mathcal{I}_{\theta}(x)=\overline{\mathcal{I}}(x)+\sum_{m=1}^{M}\theta_m\sqrt{\lambda_{m}}\mathcal{I}_{\Phi_m}(x)\]

For the computation of the new pose, note that the pose transforms are rigid transforms that belong to the manifold, and so cannot be directly obtained  as a linear combination of the pose deformation fields $\mathcal{P}_{\Phi_m}$. The generation of the pose transformation requires an exponential mapping to project the pose deformation fields from the latent space onto the manifold.

We define the exponential mapping (inverse of the log mapping in \ref{EDR}) as:
\begin{align}
&\exp\left[\sum_{m=1}^{M}\theta_m\sqrt{\lambda_{m}}\mathcal{P}_{\Phi_m}\right]=exp_{\theta}
\end{align}
with \begin{equation*}
exp_{\theta}=\underset{ h\in \textrm{SE}(3)}{\arg min} \left \Vert \overrightarrow{h\left[\sum_{m=1}^{M}\theta_m\sqrt{\lambda_{m}}\mathcal{P}_{\Phi_m}(\Gamma)\right]}-\vec{\Gamma}\right \Vert ^2
\end{equation*}
The new rigid transformation $\mathcal{P}_{\theta}$ representing the pose of the new object is then defined using the exponential mapping as:

\[\mathcal{P}_{\theta}(x)=\overline{\mathcal{P}}\circ \exp\left[\sum_{m=1}^{M}\theta_m\sqrt{\lambda_{m}}\mathcal{P}_{\Phi_m}\right](x)\]

\subsubsection{Computing new joint samples}
Similar to the training function, we compute a new joint object from the shared latent space for a given Gaussian noise $\theta$ using the sampling function $f_{\theta}$ defined as
\begin{align}\label{new joint deformation}
  f_{\theta}\colon\cup_{j=1}^{N}\Omega^j&\longrightarrow\mathbb{R}^{3}\times\mathbb{R}^{3}\times\mathbb{R} \nonumber  \\
    x&\longmapsto f_{\theta}(x)=\left(\mathcal{S}^j_{\theta}(x),\mathcal{P}^j_{\theta}(x),\mathcal{I}^j_{\theta}(x)\right), \text{ if } x\in \Omega^j
\end{align}

where $\mathcal{S}^j,\mathcal{P}^j$ and $\mathcal{I}^j$ are defined in section \ref{Computing training functions}. The new joint object is computed as

\[V_{\theta}=\{V_{\theta}^1,\dots,V_{\theta}^N\}\]
 with 
\[ V^j_{\theta}=\left\{ \left(\mathcal{P}^j_{\theta}\circ \mathcal{S}^j_{\theta}(x),\mathcal{I}^j_{\theta}(x)\right)~|~ x\in \Omega^j\right\}\]

\subsection{2D projection from 3D samples}
It maybe of interest to compute $2D$ objects from our $3D$ models in order to for example,  analyse $2D$ images such as X-ray \cite{reyneke2018review,thusini2020uncertainty}. This can be achieved in our modelling approach through a transformation that maps a $3D$ model instance  into the $2D$ image space. We define such a projection transformation $\mathcal{G}$ to be:
\[
  \mathcal{G}\colon \Theta \to \mathbb{R}^2,\quad \theta  \longmapsto \mathcal{P}(\theta )=\mathcal{G}\circ f_{\theta}
\]
where $\mathcal{G}\circ f_{\theta}$ is a $2D$ image representation of the project model deformation fields $f_{\theta}$. We thus get:
\[
\mathcal{G}\circ f_{\theta}: x\in\mathbb{R}^{2} \longmapsto I_{\theta}(x)
\]
where $I_{\theta}$ is the model's sample (image pixel) in $2D$ space.

\section{Properties of the models}\label{Properties}
As GP-based models, GMFC-GPM inherits their intrinsic properties and allows for more flexible poses through permutation.
\subsection{Intrinsic attributes}
\subsubsection{Marginalising the model}\label{model marginalization}
As in GPMMs, the DMFC-GPM approach when combined with a suitable model fitting approach allows for the estimation of missing parts a specific part of the domain our model is defined on; thanks to the marginal probability property of Gaussian processes . 

Let us assume that the DMFC-GPM is a probabilistic joint model denoted by  $\mathcal{M}(\Omega)$. The model could be marginalised over a subdomain as inherited in GPMMs, which could be features: from a specific class, from a specific structure, or from different classes. For example, the DMFC-GPM of the shoulder complex ( including scapula and humerus and clavicle) could be marginalised into three different individual object models. Furthermore, DMFC-GPMs include static multi-object models, that is to say, setting the pose parameters constant to obtain classical SSMs of articulating joints. Conversely, it also includes motion models, that is to say, setting the shape parameters constant would lead to a pose-only statistical model.  Given a subdomain $A\subset\Omega$, its marginalised models are  defined as 
 \begin{align}\label{marginalisation equation}
   \mathcal{M}(A):= \mathcal{GP}(\bar{f}, \mathcal{K})|_A.
 \end{align}
The DMFC-GPMs marginalised to shape and pose or the DMFC-GPMs with only shape and pose (that is the DMFC-GPMs with intensity constant) is termed a Dynamic Multi-Object Gaussian Process Models (DMO-GPM). We have previously published the DMO-GPMs \cite{fouefack2020dynamic}, which extend SSPMs through not only making them inherently continuous, but also by embedding a robust rigid transformation modelling metric in the approach.

Below is the formulation of the DMO-GPMs built as DMFC-GPMs with only shape-pose variability \cite{fouefack2020dynamic}. We define the DMO-GPMs as:
\begin{equation}\label{DMOGPMexperiment}
f_{\textrm{DMO}}\sim \mathcal{GP}(\bar{f}_{\textrm{DMO}}, \mathcal{K}_{\textrm{DMO}}) 
\end{equation}
where $\bar{f}_{\textrm{DMO}}$ and $\mathcal{K}_{\textrm{DMO}}$ are defined as in eq \ref{mean deformation} and \ref{kernel}, but with $f_i$ having shape ($\mathcal{S}_i$) and pose ($\mathcal{P}_i$) deformation field components only.

\subsubsection{Posterior models}\label{DMO posterrior}
Statistical models usually used in the fitting process leverage the correlation that exists within features to perform the prediction while only having partial observation of the target object. The correlation may not exist in some scenarios but one may still want to constraint these/their model to limited observations.
An alternative to correlation-based analysis is regression analysis.  Let us consider that there is no correlation within features, but we are able to observe part of the multi feature-class deformation fields, that is to say, for instance, knowing some corresponding feature of the reference domain on the observation.  
Let us consider that we have an observed joint  $O=\{ {V^1},\dots ,{V^N}\}$, with $N$ objects at various poses. Let us further assume that we know some points on the reference domain and their correspondence on the observed joint, noted as  $\{x_O, f_O,\sigma^2\}$, where $x_{O}$ are observed features on the reference domain constituting the set $\Omega_O$, and $f_{O}$ the deformation field mapping them to the target ones; $\sigma^2$ is the Gaussian noise level of the observation. We can formulate the dynamic multi feature-class Gaussian process (DMFC-GP) regression models \cite{williams2006gaussian,luthi2017gaussian} as:
\begin{equation}
   f|x_O: f_{O}\sim \mathcal{GP}(\bar{f}_O,\mathcal{K}_O) 
\end{equation}
by conditioning the mean: 
\begin{equation}
\bar{f}_{O}(x)=\bar{f}(x)+\mathcal{K}(x,x_O)^T(\mathcal{K}(x_O,x_O)+\sigma^2I)^{-1}\, (x_O-\bar{f}(x_O))\\
 ~~ 
\end{equation}
 and conditioning the kernel:
 \begin{equation}
\mathcal{K}_O(x,y)=\mathcal{K}(x,y)-\mathcal{K}_{cond}\\
\end{equation}
where 
\begin{equation}
   \mathcal{K}_{cond}=\mathcal{K}(x,x_O)^T(\mathcal{K}(x_O,x_O)+\sigma^2I)^{-1}\, \mathcal{K}(x_O,y) 
\end{equation}
with
\begin{equation}
  \mathcal{K}(x_O,x_O)=\frac{\sum_{i=1}^{n}(f_{O}(x_O)-\bar{f}_{O}(x_O))(f_{O}(x_O)-\bar{f}_{O}(x_O))^T}{n-1}
\end{equation}
 \begin{equation}
 \mathcal{K}(x_O,y)=\frac{\sum_{i=1}^{n}(f_{O}(x_O)-\bar{f}_{O}(x_O))(f(y)-\bar{f}(y))^T}{n-1}\\    
 \end{equation}
where $x_O \in \Omega_O, x,y\in \Omega$

\subsection{Pose permutation attribute}\label{permutationmodel}
The basis functions $\{\Phi_m\}$ over which the $\mathcal{GP}$ is built have pose and shape components, and it is apriori known that an object could have multiple possible poses that define its motion, relative to neighbouring objects. In practice, there are various constraints for obtaining possible poses of a given object. These constraints include limitation of patient's movement in image acquisition instruments (scanners) and high cost associated with imaging. Building a model with different motion ranges for different objects may result in spurious shape-pose correlations. To overcome this difficulty, a pose permutation-based modelling approach can be leveraged from the DMFC-GPM framework. The permutation approach consists of associating poses from other objects in training dataset to each object.

Given the training dataset deformation fields $\{(\mathcal{S}_i,\mathcal{P}_{i},\mathcal{I}_{i}), i=1,\dots n\}$, the permutation training is defined as;
$$\{(\mathcal{S}_{i},\mathcal{P}_{j},\mathcal{I}_{i}), j=1,\dots, n,i=1,\dots, n\}.$$

The resulting model may be more general, which is desired. However, the price of this generalisation may be the encoding of motion that is unrealistic for a joint (or biological) structures in the model since allowable motion is typically constrained by the shape articulations. To mitigate this drawback, the permutation defined above can be constrained using expert prior knowledge about the shape-pose relationship for a specific joint. A simple manner to to this would be to associate the pose of one joint to another one only if the shapes associated with the two joints have a certain degree of morphological similarity as determined by an appropriate metric.

\section{Experimental validation on synthetic data}\label{Experiments and Results}
\begin{figure*}[ht]
	\centering
	\includegraphics[width=0.9 \textwidth, angle =0 ]{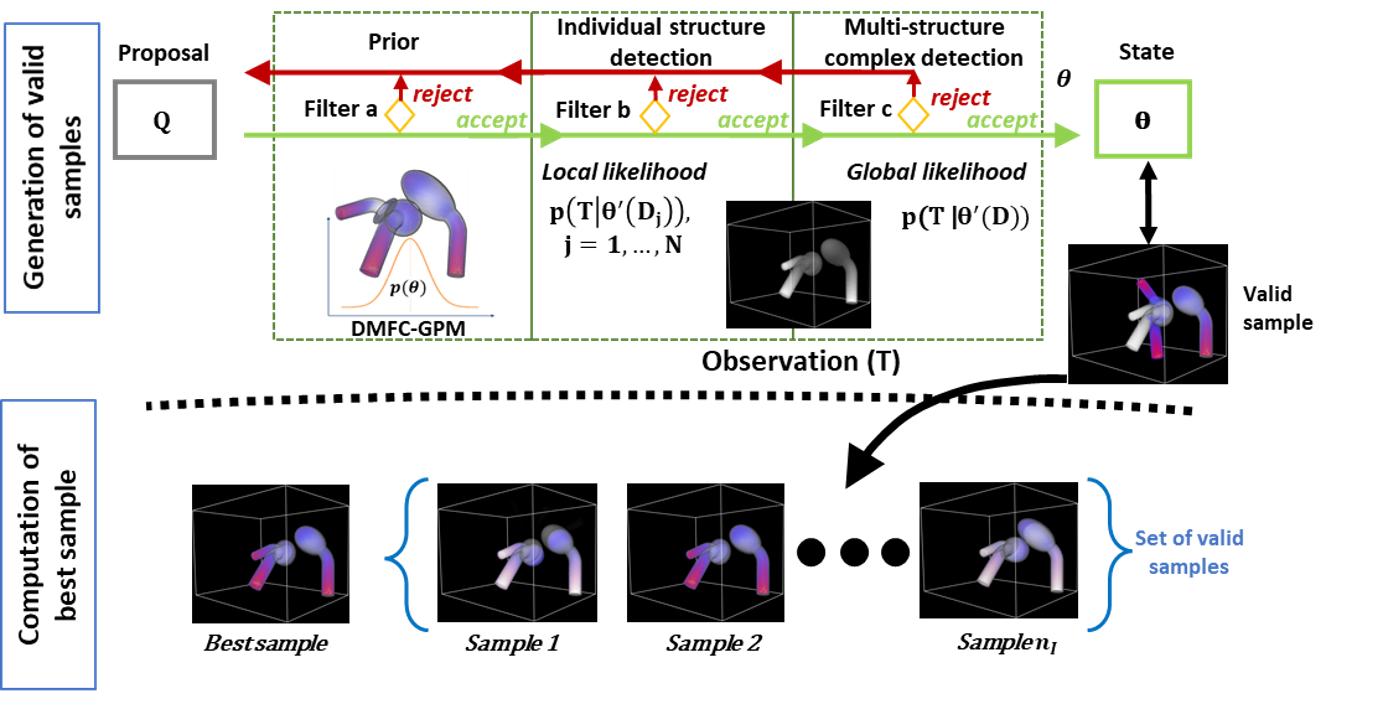}
	\caption{Adaptation of the Markov chain Monte Carlos and Metropolis Hastings methods. Top: generation of valid samples. A proposal distribution is used to propose samples. Local and global likelihoods are added to the prediction process as filters to ensure local and global similarity between the proposed sample and the observation. A sample is valid and is added to the chain if it is accepted by all filters (a, b and c), otherwise it is rejected. The process is repeated $n_I$ times to build the chain of valid samples with their corresponding probabilities. Bottom: computation of the best sample. The best sample is estimated as the one with the highest probability.} 
	\label{MCMCschmetatic}
\end{figure*}
As mentioned above, in this paper, we refer to a framework as model plus model fitting. In this section we introduce a model fitting approach to pair with our modelling approach and then we conduct experimental validation of the full framework. The experiments described were conducted in two stages: the first stage focused on validating the framework using synthetic data (Section 4). The second stage was to demonstrate the clinical application of the framework using real medical image data (section 5). 

\subsection{Fitting the model to observations}\label{DMFC-GPM fitting}
To demonstrate the utility of the DMFC-GPM framework in making predictions from observations of data, we adopt the Markov Chain Monte Carlo (MCMC) and Metropolis Hastings algorithms for the model fitting as in \cite{morel2018probabilistic,madsen2019closest,schonborn2017markov}. Markov Chain Monte Carlo methods are computational tools to perform approximate inference with intractable probabilistic models. They make use of another distribution from which one can sample, to ultimately draw samples from the true posterior distribution model (see figure \ref{MCMCschmetatic}).

Let us formulate the process of fitting using a DMFC-GPM.  Let $\mathcal{I}$ be an observation; $\mathcal{I}$ contains a joint with $N$ objects at various relative positions. The posterior model is estimated using the Bayes rule:
\begin{equation}\label{posterior}
p(\theta|\mathcal{I})=\dfrac{p(\mathcal{I}|\theta)p(\theta)}{p(\mathcal{I})}
\end{equation} 
where $p(\mathcal{I})$ is the probability of the image and is intractable; $p(\theta)$ is the model and $p(\mathcal{I}|\theta)$ the likelihood. Finding the optimal parameter in Eq. \ref{posterior} is equivalent to the following :  
\begin{equation}\label{posteriormax}
\theta^*=\underset{ \theta}{\arg max} ~p(\mathcal{I}|\theta)p(\theta)
\end{equation}
To estimate the posterior DMFC-GPM, a multi feature-class likelihood is proposed as below.

\subsubsection{Multi feature-class likelihood}
Full shape, pose and intensity estimations require a likelihood that captures within- and between structure feature-classes. We propose multi-feature class likelihoods that have global and local components. Global components ensure the prediction of all feature classes while each local component ensures the prediction of local feature classes in the global likelihood. The $j^{th}$ local likelihood $l_j$ compares the observation of the $j^{th}$ structure with its instance $\theta(\Omega^{j})$ generated from the model using independent feature-wise comparison. It can be seen as a probability distribution of possible shape, pose and intensity samples evaluated for the image data and it is defined as: 
\begin{equation}\label{single likelihood}
l_j(\theta,I)\sim p(I(\Omega^{j})|\theta(\Omega^{j}),\mu, \sigma)=\prod_{x\in \Omega^{j}}p(I(x)|\theta(x),\mu, \sigma)
\end{equation}
where $p(.,\mu,\sigma)$ is a probability density function (pdf) with the median/mean $\mu$ and the scale/variance $\sigma$.
The global likelihood $l$ compares the observation of the structures of interest in the image data with the model instance $\theta(\Omega)$ and it is defined as: 
\begin{align}\label{global likelihood}
l(\theta,I)\sim p(I(\Omega)|\theta(\Omega),\mu, \sigma)=\prod_{x\in \Omega}p(I(x)|\theta(x),\mu, \sigma)
\end{align}

\subsubsection{Obtaining the optimal parameter set $\theta$}
A sample $\theta^{'}$ is accepted as a new state $\theta$ for the $j^{th}$ object with a Metropolis decision probability $\beta_j$, and the target $I$ with a probability $\beta$.  The probabilities are defined as:

 \begin{align}\label{betaj}
 \beta_j=\min\left\{\dfrac{l_j(\theta,I)p(\theta)}{l_j(\theta^{'}, I)p(\theta^{'})}\dfrac{Q(\theta^{'}|\theta)}{Q(\theta^{'}|\theta)},1\right\}
\end{align} 

\begin{align}\label{beta}
 \beta=\min\left\{\dfrac{l(\theta, I)p(\theta)}{l(\theta^{'}, I)p(\theta^{'})}\dfrac{Q(\theta^{'}|\theta)}{Q(\theta^{'}|\theta)},1\right\}
\end{align} 
where $Q(\theta^{'}|\theta)\sim\mathcal{N}(\theta^{'}|\theta, \Sigma_{\theta})$ is the proposal generator. Finally, the proposal $Q$ is  fed  through  a sequence of  Metropolis acceptance decisions:
\begin{align}
\{\beta_j, \beta\},j=1\dots,N
\end{align}
The predicted target is the instance of the posterior model with the highest probability $\beta$ as described in figure \ref{MCMCschmetatic}.

\subsection{Description of synthetic data} \label{Description of synthetic data}
\begin{figure}[ht]
	\centering
	\includegraphics[width=0.45 \textwidth, angle =0 ]{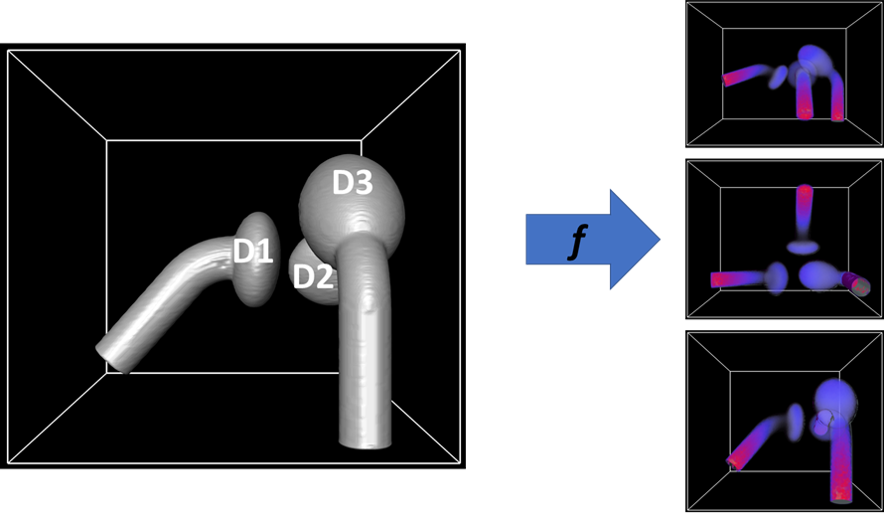}
	\caption{$3D$ image render from a joint lollipop. Left: joint lollipops with three objects. Right: cross view of the $3D$ image rendered. } 
	\label{lollioprender}
\end{figure}

We evaluated the DMFC-GPM framework on synthetic data designed with controlled shape and motion variation. The use of synthetic data avoided dealing with the complex variation in shape, pose and intensity from real images for which ground truths would not be available. Our synthetic data consisted of surface mesh data of  "lollipops" as defined in \cite{gee2014systematic}. We created artificial joint-like structures using three lollipops per sample. Each joint was composed of three lollipops with major axes for corresponding pairs of lollipops of $r_1$, $r_2$ and $r_3$, for object 1, object 2 and object 3, respectively. The span of the dataset of joints was created by varying $r_1$ and $r_2$ as $\{(r_1,r_2,r_3)=(r, 31-r,17-r),i=1,\dots,15\}$ creating a shape correlation between them. For joint motion,  for each joint generated above, we rotated the second lollipop (object 2) relative to the first one (object 1) and the third relatively to object 2 using the Euler's angle convention for describing rigid transformations $(\varphi,\theta, \psi)$. The second lollipop was moved in the $yz$-plane  by four angles ($\theta=\frac{1}{5}\pi,\frac{2}{5}\pi,\frac{3}{5}\pi,\frac{4}{5}\pi$). The third object was moved in the $yz$-plane  by four angles ($\theta=\frac{1}{2}\pi,\frac{1}{3}\pi,\frac{2}{9}\pi,\frac{1}{9}\pi$) defining a Euclidean motion that is highly correlated  with that of the second object. To get joint lollipops , we rendered joint lollipops.  Image domains were defined using bounding box of the joint and  pixel values were assigned to  point in the domain using the following function: 
 \begin{align}\label{imagerenderingfunction}
f(x)= \begin{cases}
\|x-x_1\|, \text{ if } x\in D_1\\
\|x-x_2\|,\text{ if } x\in D_2\\
\|x-x_3\|,\text{ if } x\in D_3\\
0, otherwise.
\end{cases}
\end{align}
where $D_1, D_2, D_3$ are the shape domain of the first, the second and the third object, respectively. Figure \ref{lollioprender} shows an example of image rendering from a lollipop joint.

The lollipop surface mesh data described above were in correspondence, however, to establish intensity correspondence, we employed nearest-neighbour interpolation of the shape correspondence transformations. The interpolated transformations were applied to tetrahedral meshes of the references for each structure to obtain in-correspondence tetrahedral meshes. These tetrahedral meshes were used to sample intensities from the image volumes, thus establishing shape, pose and intensity correspondences across the training data.

\subsection{Model sampling and evaluating correlation}\label{validation with synthetic}
Our first experiment evaluated the ability of the DMFC-GPM to explain within and between feature-class (including intensity) correlations. Samples of the DMFC-GPM built from the lollipop data were randomly generated. Between object shape correlation, shape-intensity correlation, as well as between object pose correlation of generated samples were evaluated. The results were compared to the known correlations from the training examples.
 
 The results show that the DMFC-GPM framework adequately explained  the correlation from the training data with the correlation percentage being approximately the same in both cases (training examples and model samples). Samples from $-2$ and $+2$ standard deviations around the mean were generated for the first and the second modes of variation of the model. The first principal geodesic \footnote{We refer to the principal geodesic instead of the principal component. Indeed, DMFC-GPMs deal with the variability of joint features that are linear (shape and intensity) and non-linear (pose), so the PGA is used for the model computation.} (PG) accounted for $86\%$ of the total variation and explained the shape and intensity variation encountered in the training data. The second PG accounted for  $9.3\%$ of the total variation, explaining joint motion (pose variation). This was expected as there was a correlation between shape and intensity distribution, hence, both accounted for high variability across the training data.  

\begin{figure*}[ht]
	\centering
	\includegraphics[width=0.8 \textwidth, angle =0 ]{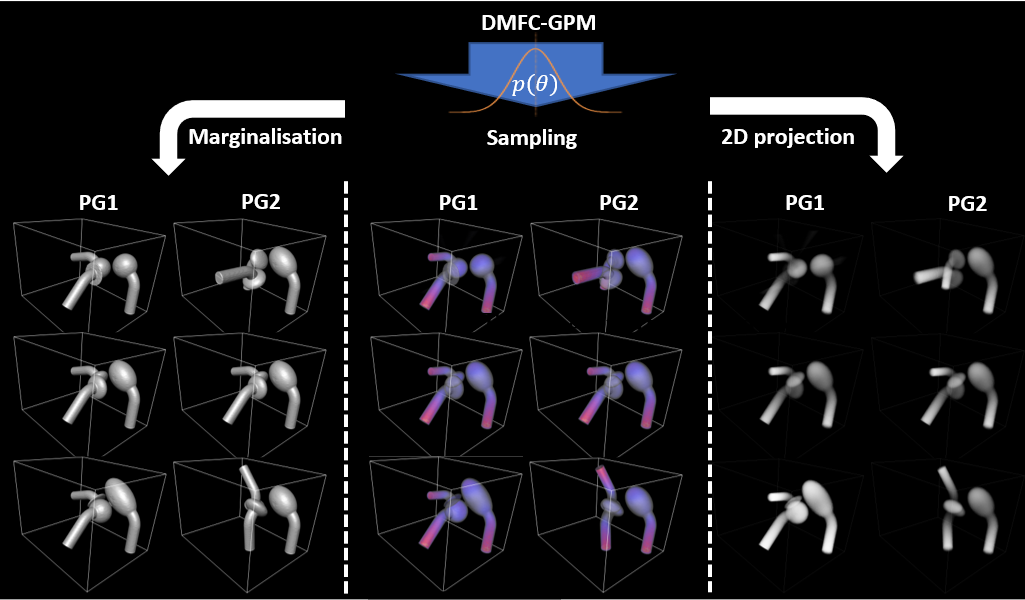}
	\caption[Sampling of the lollipop DMFC-GPM]{Sampling of the lollipop DMFC-GPM. Top: the DMF-GPMs ($p(\theta)$ is the probabilistic formulation of the model). Bottom: shape, shape-intensity and image modality sampling. From left to right: Sampling from the first and the second PG of shape joint, joints with shape and intensity , $3D$ image rendered and DRRs. Samples from $-3$ and $+3$ std around the mean.} 
	\label{DMFCSampling}
\end{figure*}

Next, joint shape samples were generated along with joint shape and intensity as observed in figure \ref{DMFCSampling}. A volume renderer was also added in the framework to generate $3D$ images along with a $3D$ to $2D$ projection method (digitally reconstructed radiographs - DRRs) as observed in figure \ref{DMFCSampling}. The projection was done using direct volume ray-casting, to simulate radiographic images \cite{max1995optical}.

\begin {table*}[!htbp]
\caption[Correlation coefficient between and within feature classes] {Correlation coefficients between and within feature classes. Comparison of the correlation between the model samples and the training examples. $d_1$, $d_2$, $d_3$ are the intensity information for the first, the second and third object, respectively. $\theta_2$ and $\theta_3$ are the motion parameters of the second and the third objects, respectively.}
\label{Correlation coefficient} 
 \begin{center}
    \begin{tabular}{|p{3.5cm}| p{1.5cm} |p{1.5cm} |p{1.5cm} |p{1.5cm}| p{1.5cm}| p{1.5cm}|}
    \hline
     & $r_1$ vs $d_1$ & $r_2$ vs $d_2$ & $r_3$ vs $d_3$ & $\theta_2$ vs $\theta_3$& $r_1$ vs $r_2$ & $r_2$ vs $r_3$\\
    \hline
    Training & $46\%$ & $92\%$& $96\%$& $60\%$&$99\%$& $93\%$\\
        \hline
   \textbf{DMFC-GPM} & $\textbf{56\%}$& $\textbf{92\%}$& $\textbf{93\%}$& $\textbf{53\%}$&$\textbf{98\%}$& $\textbf{91\%}$\\
    \hline
\end{tabular}
\end{center}
\end{table*}

To evaluate the cross-correlation between different feature classes and objects, $100$ samples were randomly generated from DMFC-GPM of the lollipop data. Since the intensity features in the lollipop structures were assigned as a function of shape ($r_1$, $r_2$, $r_3$), the intensity quantities ($d_1$,$d_2$, $d_3$) were computed using the same points used to compute shape quantities. For the pose quantities ($\theta_1,\theta_2,\theta_3$), the first Euler parameters were used which describe the motion simulated in the training dataset. Between object shape correlations ($r_1 vs. r_2$ and $r_2 vs. r_3$), shape-intensity correlations ($r_1 vs. d_1$, $r_2 vs. d_2$ and $r_3 vs. d_3$), as well as between object pose correlations  ($\theta_2 vs. \theta_3$) of the DMFC-GPM samples, were evaluated. The correlation coefficients from the DMFC-GPM's samples were compared to the those from the training examples. 

The results are shown in table \ref{Correlation coefficient}. It can be seen that the DMFC-GPM adequately explains the correlations from the training data with the correlation percentage being approximately the same in both cases. The slight difference between the shape and intensity for the first object could be attributed to random sampling of the model.

\subsection{Comparison with other models }\label{Shape, pose and intensity modelling validation}
We compared the DMFC-GPM framework with other methods for shape pose and intensity prediction from images. In all comparisons we used common synthetic data and a common model-fitting approach (MCMC sample-based fitting).

\textbf{\textit{Encoding prescribed motion}}: 
  We compared the performance of the EDR (from our model), the SR, and point distribution model (PDM) in explaining the transforms describing degrees of freedom (DOF) of a prescribed motion. For the  comparison, the DMO-GPM was used (recall that the DMO-GPM is a marginalised DMFC-GPM over shape and pose only as described in section \ref{model marginalization}). To implement DMO-GPM with SR, rigid transformations within the DMO-GPM setting were represented by Euler's parameters, that is, rotations were represented by Euler's angles and the translation by $3D$ Euclidean vectors.

\begin{figure*}[ht]
	\centering
	\includegraphics[width=1.0 \textwidth, angle =0 ]{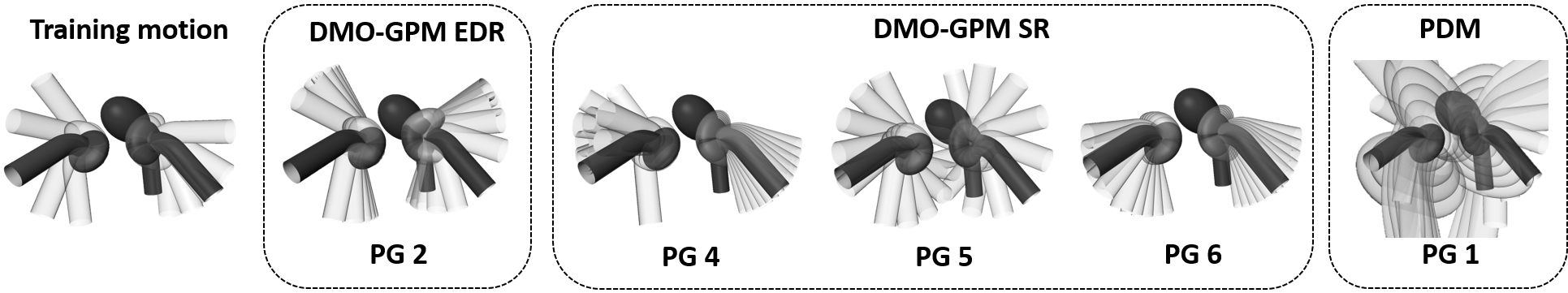}
	\caption[Pose comparison with sampling]{Comparison of pose variations with the training data. From the left to the right: Pose in the training data, pose of the DMFC-GPM with EDR, pose of the DMO-GPM with SR (pose regression) and pose with PDM.} 
	\label{posecomparisonDMO_EDRvsSR}
\end{figure*}

  Figure \ref{posecomparisonDMO_EDRvsSR}, from the left to the right, shows the DOF of the training data, the pose variation of DMO-GPM with EDR, the pose variation of the DMO-GPM with SR and the one with PDM. Our model captures the training motion with one PG while the one with SR needs up to $3$ PGs to explain the motion that is in one plane in the training data. As expected PDM led to unrealistic shape variation.
 
 \textbf{\textit{Specificity and generality}}:
The specificity and generalisation of the lollipop DMFC-GPM with EDR was compared to DMFC-GPM with SR. For testing, a synthetically created dynamic $3D$ image dataset (joint with shape and intensity classes at various poses) was used. The images that were included in the training dataset were predicted with both models (EDR-based DMFC-GPM and SR-based DMFC-GPM) to compare the specificity performance. The images had six poses, meaning each of the six images displayed object 2 and object 3 at specific poses. 

\begin{figure}[ht]
	\centering
	\includegraphics[width=0.5 \textwidth, angle =0 ]{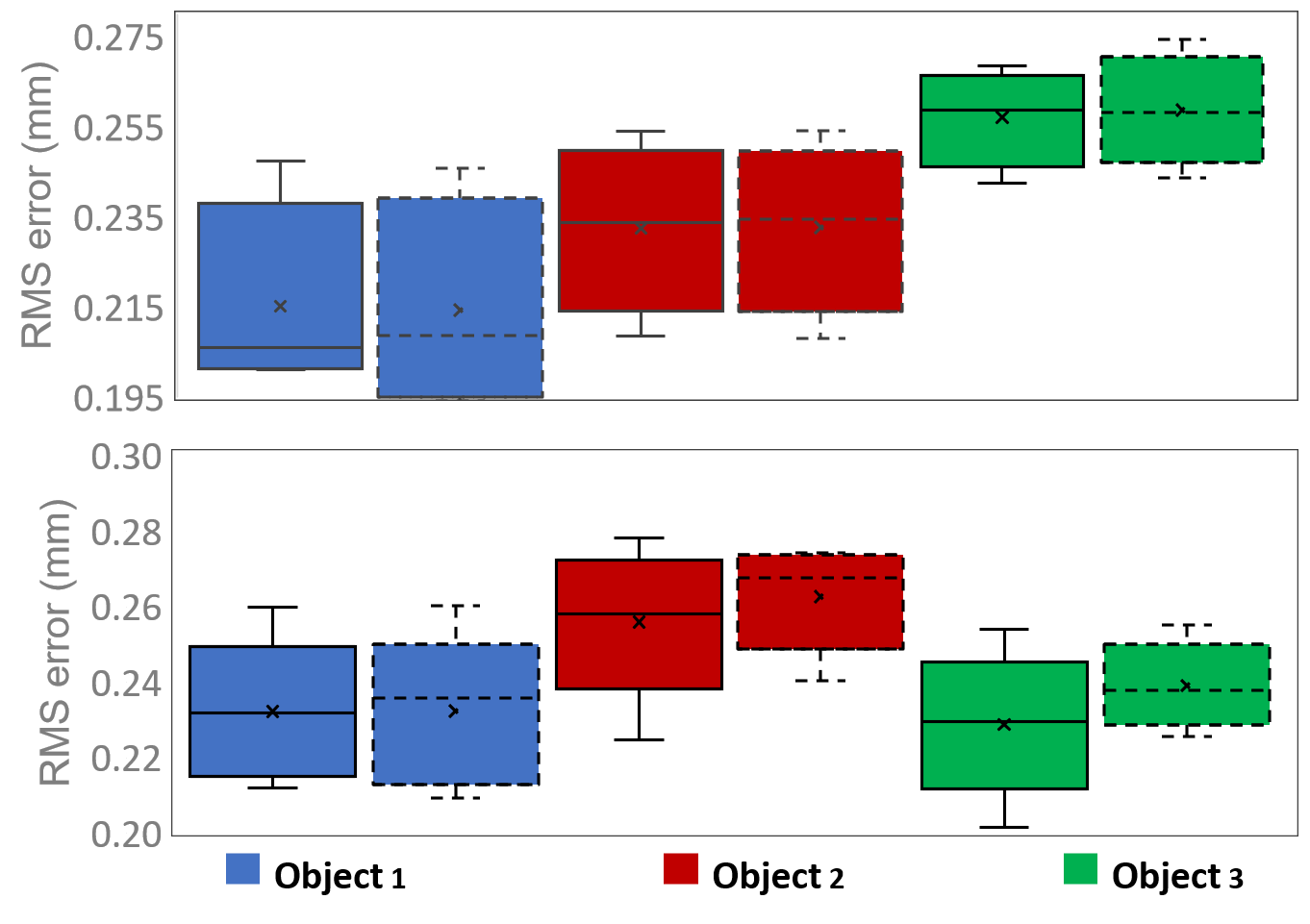}
	\caption[Generalisation and specificity of the lollipop DMFC-GPM]{Generalisation (top) and specificity (bottom) of the DMFC-GPM with EDR and the  DMFC-GPM  with SR. The intensity based RMS errors of the prediction using EDR (solid line) and the one using SR (dashed line) for first (blue), second (red) and third (green)  objects.} 
	\label{EDR vs SR}
\end{figure}

 Figure \ref{EDR vs SR} (bottom) shows the box plot of the intensity-based RMS errors of the predicted and observed intensity distribution, where the model with EDR outperforms the one with SR. For generalisation performance, a dynamic $3D$ image dataset with joint poses not included in the training was generated and predicted using both models. Figure \ref{EDR vs SR} (top) shows the RMS errors and for each object, the model with  EDR outperforms the one with SR. 

 \textbf{\textit{Segmentation}}:
 The ability of the DMFC-GPM to segment objects from image volumes was compared to that of an ASM. Gaussian process morphable model-based ASM and DMFC-GPM were used to predict an individual object and the results were compared. A test dataset was created, consisting of $6$ lollipop joints, for each joint. A $3D$ image was rendered with a missing part on the object to be predicted, leading to a set of images with missing structures. The DMFC-GPM and individual ASM were fitted onto the test $3D$ images to predict target objects including the missing parts. The RMS errors of the shapes predicted using  DMFC-GPM and those predicted using GPMM-based ASM were compared. To implement the GPMM-based ASM, first, the GPMM-based SSM was developed from in-correspondence mesh surfaces for each object. Second, objects were aligned with their corresponding $3D$ image volumes and the intensity profile was generated at each vertex of the mesh surface. Finally, the average intensity profile was computed. For each individual object, the SSM and the average intensity profile were simultaneously used within the fitting setting for the prediction.
  
\begin{figure}[ht]
	\centering
	\includegraphics[width=0.45 \textwidth, angle =0 ]{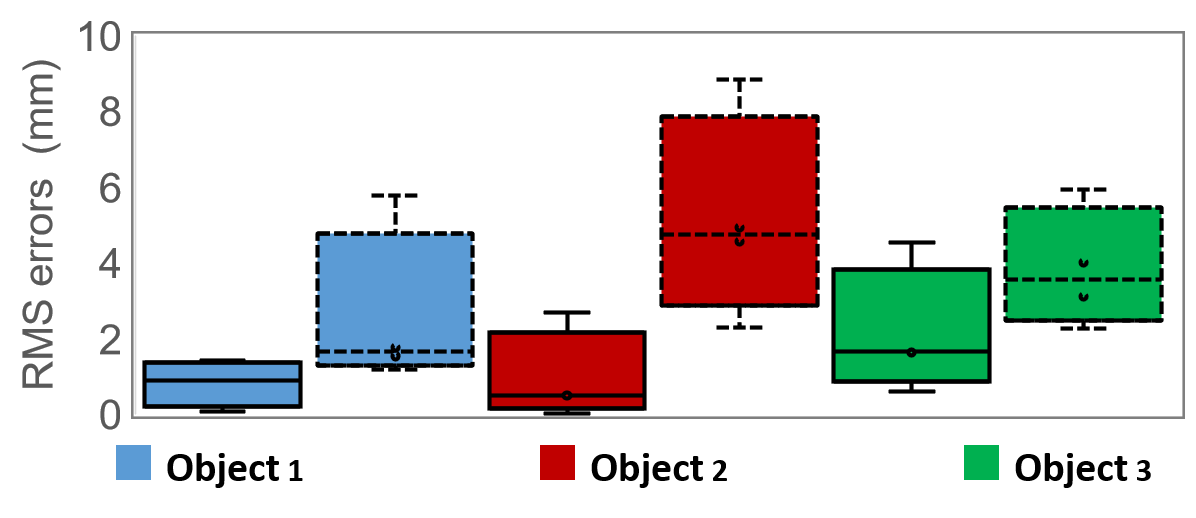}
	\caption[RMS distance errors of lollipop DMFC-GPM]{RMS distance errors for the prediction from $3D$ images using  DMF-GPMs (solid line) and ASMs (dashed line). The error for the DFCM-GPM and the ASM for the first (blue), second (red) and third (green) objects.} 
	\label{DMFCvsASM_RMS}
\end{figure}

  Figure \ref{DMFCvsASM_RMS} shows surface to surface RMS distance errors of the prediction with DMFC-GPM and ASM.
  It can be seen that the prediction errors of the DMFC-GPM framework (solid line) are smaller compared to the GPMM-based ASM ones (dashed line) indicating that the DMFC-GPM outperformed ASM on this prediction task.
  The performance gains from using the DMFC-GPM could be attributed to the additional knowledge (shape information on the observation) obtained from the correlation between objects embedded in that type of model. 
  
\section{Example application on shoulder data}\label{applications}

\subsection{Description of shoulder data}\label{shoulder data}
For the demonstration of the framework on real data, two case studies were performed. The first experiment was on shoulder joint image data (glenohumeral joint) (SITE 1) with a high range of simulated motion as described in \cite{fouefack2019statistical}. The second experiment used data of the shoulder complex (glenohumeral joint and acromioclavicular joint) (SITE 2) with real but small range of motion. The data from SITE 1 consisted of $3D$ mesh surfaces segmented from CT images of the bilateral shoulders of $18$ fresh cadavers. The cadaveric specimens were collected from the Division of Clinical Anatomy and Biological Anthropology, Faculty of Health Sciences, University of Cape Town, South Africa. Institutional ethics approval was granted for the study (HREC 283/2020). SITE 2 data consisted of $12$ bilateral cadaveric shoulder complex CT images (total of 24 shoulders) obtained from the SICAS Medical Repository (SICAS: http://www.si-cas.com/). These data are publicly available for research use. The voxel size of the images ranged from  $0.97 \times 0.9 \times 0.5$ to $1.27\times1.27\times0.8~ mm$. These image data were obtained from cadavers and each image data volume included bilateral images of clavicle, humerus, and scapula. 

Shape correspondence was established using the model-based fitting approach with free form deformation models constructed as GPMMs through rigid and non-rigid registration as described in \cite{fouefack2020dynamicarxiv}. The relative positions of the objects were recovered by applying the inverse of the rigid transformation to the objects resulting from the registration. Intensity correspondence was established as described in section \ref{Description of synthetic data}.
\begin{figure*}[t]
	\centering
	\includegraphics[width=0.95 \textwidth, angle =0 ]{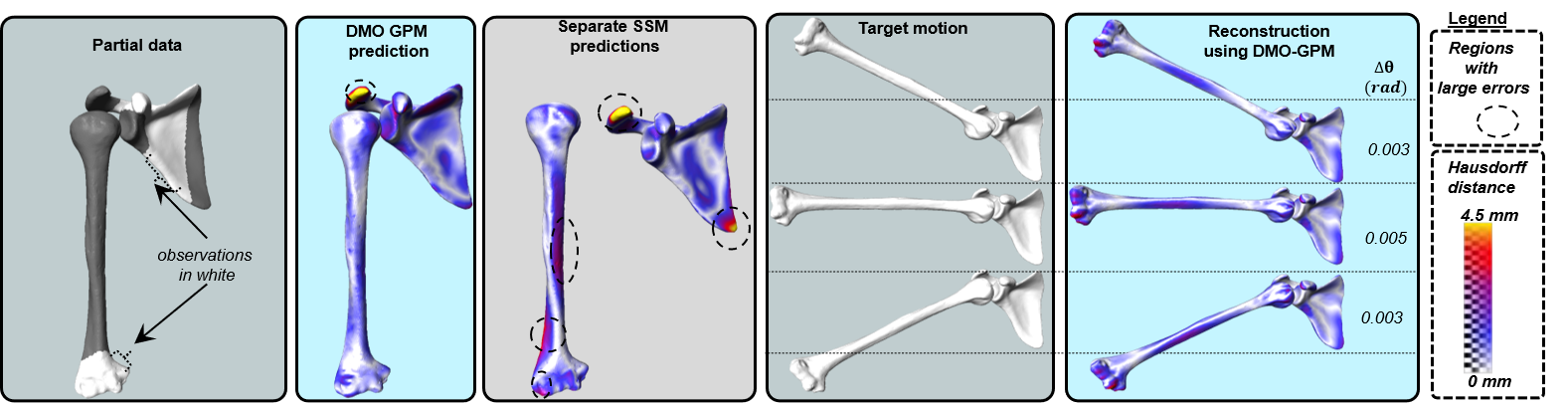}
	\caption{Reconstruction of target shoulders. From left to right: partial shoulder to be reconstructed, reconstruction using DMO-GPM marginalized shape, reconstruction using PDM, target motion to be reconstructed, prediction of DMO-GPM marginalized pose at multiple poses.} 
	\label{fitmotionDMO}
\end{figure*}
\subsection{Shape and pose analysis}
We assessed the shoulder DMFC-GPM for predicting pre-morbid shape and also encoding a prescribed motion using the DMO-GPM sub model. We predicted the simulated abduction motion described in \cite{fouefack2019statistical} using the data from SITE 1. The scapula and the humerus were predicted with the humerus at different poses relative to the scapula. Five poses were generated within the motion range $(0,\frac{3}{10}\pi, \frac{1}{2}\pi,\frac{7}{10}\pi,\frac{4}{5}\pi)$ which simulated the abduction motion in the training data-set. Furthermore, distal humeral and scapula blade fractures were simulated, and both DMO-GPM and individual bone PDMs were used to reconstruct fractured bones. The prediction was done using DMO-GPM with a comparator method using a PDM. 

The reconstructed shoulder with humerus and scapula at different relative poses using the DMO-GPM are shown in figure \ref{fitmotionDMO}. The poses were predicted with the average of $0.003 ~rad$ and the average Hausdorff distance shape error was $3.1\pm 0.4 ~mm$. The reconstructed shoulder fracture is also shown in figure \ref{fitmotionDMO} for both DMO-GPM and individual PDMs. The DMO-GPM not only preserves the relative orientation of the bone but also performs better pre-morbid shape prediction.

\subsection{Shape, pose and intensity analysis}\label{Shape, pose and intensity analysis}
To evaluate the ability of the DMFC-GPM to analyse the shape, pose and intensity of medical images. A DMFC-GPM was constructed from SITE 2 shoulder complex data.

\begin{figure}[ht]
	\centering
	\includegraphics[width=0.4 \textwidth, angle =0 ]{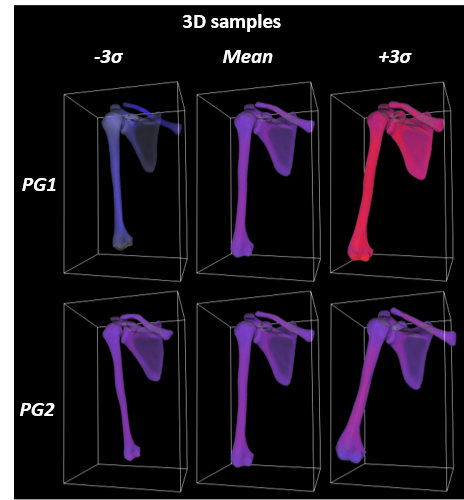}
	\caption[DMFC-GPM of the shoulder sampling]{Sampling from the DMFC-GPM of the shoulder. Left: Top to bottom: Samples of the first and second PGs of the volumetric joint with an intensity of about $-3$ to $+3$ standard deviation around the mean. 
	} 
	\label{ShoulderDMFCsampling}
\end{figure}

Figure \ref{ShoulderDMFCsampling} shows the shape and intensity samples of the shoulder complexes generated along the first and second PGs. The samples were generated from $-3$ to $+3$ standard deviations from the mean. Changes in shape, pose and intensity can be observed,  where the colours indicate the intensity values in Hounsfield units (HU). For PG $1$, the intensity at the humeral shaft changes from dark red ($-3 \sigma$ ) to light red ($+3 \sigma$) as well as the angle between the medial border of the scapula and the humerus increases from $-3 \sigma$ to $+3 \sigma$. For the same PG, the width of the humeral shaft and the subscapular fossa increases from $-3 \sigma$ to $+3 \sigma$. For PG $2$, the clavicle seems to change its orientation relative to the scapula; the deltoid attachment area becomes more curved from $+3 \sigma$ to $-3 \sigma$ about the mean.   

To automatically fit the DMFC-GPM onto a CT image data for the prediction, a test dataset of $6$ CT image volumes was used. Each shoulder complex was predicted from a CT image in the test dataset. The observations were full CT image volumes, that is, all the  slides of the CT data. The RMS errors were computed between the predicted intensity and the original joints. The errors were computed in $HU$. 

\begin{figure*}[ht]
	\centering
	\includegraphics[width=1 \textwidth, angle =0 ]{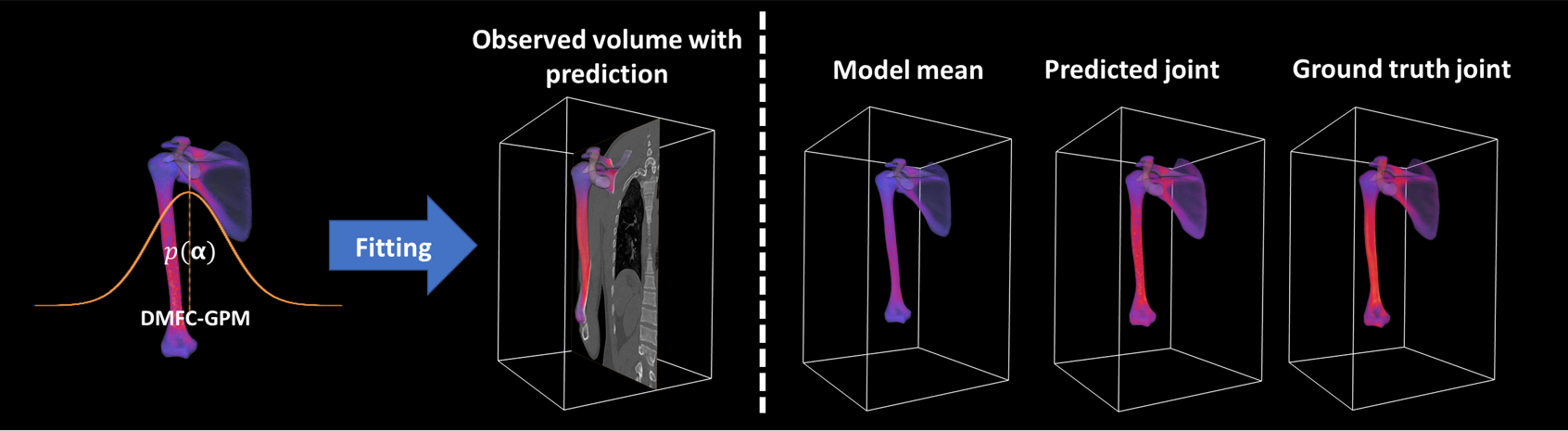}
	\caption{Shoulder prediction from CT images. Left: DMFC-GPM used for prediction from CT volume through the MCMC fitting. Right: from left to right, the mean of the model used as the initial sample for the fitting process, the predicted shape, pose and intensity profile of the joint and the ground truth.} 
	\label{ShoulderDMFCfitting}
\end{figure*}

Table \ref{prediction from CT} shows the average errors, in $HU$, of the prediction from all CT images in the test dataset. The errors were the normalised RMS similarity measures between the best sample intensity sampled from the model and the observed shoulder intensity. The prediction errors were relatively small. The average error was $0.053 ~HU$ for the clavicle, $0.065 ~HU$ for humerus and $0.040 ~HU$ for the scapula. The relative small scale of these errors can also be observed in figure \ref{ShoulderDMFCfitting}, where the best the prediction is shown in the $3D$ case. 

\begin{table}[ht]
\caption[]{Results of the prediction of the shoulder joint from the observation of CT volumes in the test dataset. The average RMS errors between the predicted intensity profile and those from the original image data of the clavicle, humerus and scapula.} 
\label{prediction from CT} 
 \begin{center}
\begin{tabular}{|c|c|c|c|}
\hline
&\multicolumn{3}{|c|}{Average RMS errors (HU)}\\
\cline{2-4}
Image   & & & \\
modality&Clavicle & Scapula& Humerus\\
\hline
 CT & & & \\
 volume& $0.053\pm 0.017$ & $0.065\pm 0.026$&$0.040\pm 0.041$\\ 
\hline
\end{tabular}
\end{center}
\end{table}

\section{Discussion}\label{Discussion}


Mathematical modelling of medical image features is a key step in analysing the image content for clinical diagnostic or decision making capabilities. While shape, pose, and intensity are considered relevant features to model, combining these features in a single representative and robust model, and furthermore for multiple articulating objects has been elusive in the literature. To this extent, this study is the first to derive a continuous multi-(feature) class statistical latent space that incorporates the combined shape, pose and intensity variations of multiple objects of interest within medical imaging data sets. The proposed DMFC-GPM framework unifies individual feature-class and single-object modelling constructs while resolving the associated problems with each feature-class. Furthermore, we present rigorous experiments on controlled synthetic imaging data as well as real image data to validate the approach as well provide an illustrative use case. This study illustrates how the proposed approach can be considered as an integration of all other single- or multi-feature modelling approaches and how it can be quickly customised to derive a specific combination of feature-class models. Finally, this study represents a new modelling paradigm which is robust, accurate, accessible, and has potential applications in a multitude of scenarios (biomechanics, clinical decision making, image processing).


Single object models are ubiquitous in the literature with proven success in segmentation tasks. However, analysing multiple objects in an image using multiple single object models may lead to large errors and uncertainties, especially around organ boundaries \cite{nakao2021statistical}. Thus, the obvious question is, are multi-object models better at multi-object segmentation when compared with single-object ones? In this paper, we answer this question through comparing the efficacy of multi-object models and single object models in lollipop object reconstruction tasks (section \ref{Shape, pose and intensity modelling validation}). Our findings suggest that indeed, a unified multi-object modelling approach is much better suited for such segmentation scenarios. One benefit of the DMFC-GPM is the approach maintains the statistics of single object variation such as SSMs \cite{cootes1992training,luthi2017gaussian}, SAMs  \cite{cootes1995active,cootes1998active} or SSPMs \cite{smoger2015statistical,klinder2008spine,fitzpatrick2011development,agrawal2020combined} and these can be retrieved on demand without any need to retrain the model. An additional benefit is that where strong correlations between different objects exist, these can be leveraged to make a prediction of a targeted object with partial or no observation of the target in the image. The uncertainty of the prediction can be calculated from the posterior of the DMFC-GPM, given observations of the other modelled objects.

Multi-object shape models are scarce in the literature.  Attempts made so far for medical image segmentation and registration include: Multi-object models with pose variation based on linear descriptions \cite{rigaud2018statistical,klinder2008spine} for segmentation of joint bones with small pose amplitude (such as vertebrae). However, these methods model the pose transformations as elements of a Euclidean vector space (linear space), in contrast to the DMFC-GPM which treats them as elements of the non-linear manifold consistent with reality. Multi-object models based on the combination of separate SSMs and SPMs \cite{chen2014automatic,yang2015automatic} have also been proposed for simultaneous  segmentation of multiple bones. However, the use of two separate shape and pose models leads to a process where the predicted shape may not conform to the anatomical constraints of the joint. The DMFC-GPMs overcome this problem by computing a shared latent space of shape and pose, which allows joint motions to be influenced by the shape of their articular geometries, thus adhering to the anatomical joint structure. Other alternative multi-object models have been based on the SR \cite{nakao2021statistical,anas2014statistical}. However, they are limited in their ability to model the rigid transformations of non-compact objects (such as the humerus). In DMFC-GPMs, this drawback is solved by using the EDR to model the pose, providing more information about the object's displacement. This makes it robust when capturing the pose descriptors of the object regardless of the object's geometry (compactness).


The notion of correlations between features has to be carefully understood from a mathematical as well as physiological perspective. Mathematically, the DMFC-GPM embeds shape, pose, and intensity as features into a combined low dimensional lower space under the assumption that a quantifiable correlation exists among them. For shape and intensity, the literature has reported that such correlation may exist \cite{cootes1998active}. Next, Wolff's law implies that form follows function which at least for articulating joints means that there exists some relationship between shape (of the bones forming a joint) and permissible poses for that shape. However, there does not seem to be any evidence supporting intensity and pose correlation. A limitation of our approach is that transitively this relationship is implied in our mathematical formulation despite not having any physiological meaning. We consider this relationship as a mere artefact of combining the feature spaces and we do not intend to use or explain this relationship within the scope of this study. 

On the other hand, combining multiple feature classes has proven advantageous in exploiting the correlation that can exist between objects and within feature classes to improve the accuracy of the prediction compared to methods that combine separate models \cite{cootes1995active,delingette2021}. However, there needs to be careful consideration to reduce spurious correlations if the training dataset is not sufficiently representative.   Indeed, incorporating multiple feature classes into the model requires a great deal of diversity (large variability in terms of shape, pose and intensity) in the training dataset in order to capture a true correlation.  For example, having an image of a patient in only one specific pose in the training dataset can lead the model to understand that patient-specific shape can only exist in that imaged position, which is not true. A realistic correlation between shape and pose may not be captured if the images of the patients included in the study were not acquired at multiple poses. That being said, the permutation property of the DMFC-GPM (section \ref{permutationmodel}) makes it possible to increase the variability of the pose within the model as discussed before.


In the previous section of this paper (section \ref{applications}), the application of the proposed framework to the prediction of shoulder joint shape, pose and intensity was presented as a use case. While the abduction-like pose variations were simulated in the  training dataset (SITE 1) to have a large pose amplitude, it represented a typical clinical scenarios in the real world. The clinical scenario relates to the estimation of a pre-morbid shape (and pose) from a partial observation of the joint for shape (and pose) feature. With regard to studies that focus on solving similar problem, the prediction of pre-morbid shoulder bone shape using SSMs has had some success in the literature \cite{abler2018statistical,pitocchi2020integration,salhi2020statistical}. The main drawback of these studies is that of maintaining anatomical joint space, as the relative position between the scapula and humerus is not taken into account when using single-object shape models. Furthermore, single-object SSMs are not able to leverage any shape correlation that may exist between adjacent bones to improve the prediction accuracy. In spite of the limited size of the training data used in the study, our DMFC-GPM framework illustrated that a shoulder joint model incorporating shape and pose features could provide additional knowledge for the estimation of the shoulder bone shapes (scapula and humerus) while considering their spatial orientations. This led to better reconstruction of the bone shapes compared to those from individual bone SSMs, while simultaneously allowing for realistic prediction of the relative pose between the bones and thus maintaining anatomical joint space of the shoulder joint. The latter is not achievable with single object SSMs and thus is a novel attribute of our framework.   

It is important to note that the volumetric correspondence (from which the intensity match is defined) in the experiment (section \ref{Description of synthetic data}) is induced from the surface match by nearest neighbour interpolation. Such an approach does not guarantee the entire flow of the intensity domain, which may reduce the ability of the model to cover the space of possible shapes and intensities within the population. The alternative would be to use the posterior model in the volumetric domain computed from the surface correspondence deformation fields to ensure the full flow of intensities.  However, the study of correspondence approaches was not the focus of this paper. Furthermore, although this model is easier to interpret due to the mechanistic relationship between the behaviour and the model parameters, its use for inference can be slow compared to deep learning approaches. However, the DMFC-GPM can be used as a decoder in the deep auto-encoder framework \cite{tewari2018high}, which leads to a more interpretative and faster model in the inference process.

Studies have reported the use of statistical model-based methods for $3D$ reconstruction from $2D$ images \cite{klima2016intensity,yao2001construction,ehlke2013fast}; and by using the same techniques, but with intensity recovery based on the associated pose and region \cite{fotsin2019shape}. However, all of the above methods work by combining a secondary and separate intensity distribution model along with the shape and pose models. The DMFC-GPM by virtue of its construct eliminates the need of using a separate, secondary intensity model. The use of DMFC-GPM may allow the prediction of the joint shape at any pose and intensity profile of the joint from the $3D$ observation volume and associated $2D$ projection, as illustrated in figure \ref{DMFCSampling} using synthetic data. The inherent continuity of the DMFC-GPM's domain offers the possibility to customise the resolution of the intensity profile, thus avoiding computational costs that can occur with previous methods \cite{klima2016intensity}. However, illustration of this application in a real world scenario is future work and out of scope for the current study.


It should also be important to mention that our efforts in this proposed work were towards developing an approach for encoding variability in morphology and features that relate to its understanding as acquired in the medical imaging process. Within this scope we did not consider the open ended problems such as high dimensionality low sample size issue \cite{hall2005geometric} or image harmonisation that exist and affect our modelling framework. However, these are not unique to our approach and impact other approaches including SSMs, ASMs, SSPMs, SPMs, and even deep learning approaches. Careful consideration is required for applying such approaches for specific applications.

\section{Conclusion}\label{Conclusion}
In this paper we proposed and evaluated a novel approach for combining statistical representations of multiple feature classes, namely shape and pose and intensity features from given medical image data sets. The formulation employs interpolated deformation fields for shape, pose and intensity feature classes that incorporate the natural representation of the objects in terms of smoothness, through a continuous dynamic multi-feature latent space. This approach provides a robust and principled way of building data-driven generative models for medical image analysis. Combining the approach with a suitable fitting algorithm provides a manner to use them in predicting individual or multiple objects, as well as single or combined feature classes. A new shape-pose modelling approach DMO-GPMs that extends GPMMs \cite{luthi2013unified}, SSPMs \cite{bossa2007multi} and SPMs \cite{moreau2017new} was also presented as an attribute of the DMFC-GPMs.  Application of the proposed DMFC-GPM framework was illustrated in shoulder joint as a holistic analysis of joints in medical images which is relevant in clinical cases. We have developed a tutorial$^{\ref{tutorial}}$ to show how to use the framework as well as to help understand the theoretical concepts used.

\bibliography{mybiblio.bib}{}
\bibliographystyle{IEEEtran}
\end{document}